\documentclass[fleqn,usenatbib]{mnras}

\usepackage[T1]{fontenc}

\DeclareRobustCommand{\VAN}[3]{#2}
\let\VANthebibliography\thebibliography
\def\thebibliography{\DeclareRobustCommand{\VAN}[3]{##3}\VANthebibliography}

\usepackage{graphicx}	% Including Figurefiles
\usepackage{amsmath}	% Advanced maths commands
\usepackage{amssymb}	% Extra maths symbols
\usepackage{caption}
\usepackage{subcaption}
\usepackage{newtxtext,newtxmath}
\usepackage{cleveref}

\usepackage{cite}
\title[An ML based FAP method]{The verification of periodicity with the use of recurrent neural networks}

\author[N. Miller et al.]{
N.~Miller$^{1}$\thanks{E-mail: n.miller4@herts.ac.uk},
P.\,W.~Lucas$^{1}$,
Y.~Sun$^{2}$,
Z.~Guo$^{3,4,5,1}$,
W.\,J.~Cooper$^{1,6}$,
C.~Morris$^{1}$
\\
$^{1}$Centre for Astrophysics Research, University of Hertfordshire, College Lane, Hatfield, Hertfordshire, AL10 9AB, UK\\
$^{2}$Centre for Computer Science and Robotics Research, University of Hertfordshire, College Lane, Hatfield, Hertfordshire, AL10 9AB, UK\\
$^{3}$Instituto de F{\'i}sica y Astronom{\'i}a, Universidad de Valpara{\'i}so, ave. Gran Breta{\~n}a, 1111, Casilla 5030, Valpara{\'i}so, CL\\
$^{4}$N\'ucleo Milenio de Formaci\'on Planetaria (NPF), ave. Gran Breta{\~n}a, 1111, Casilla 5030, Valpara{\'i}so, CL\\
$^{5}$Departamento de F{\'i}sica, Universidad Tecnic{\'a} Federico Santa Mar{\'i}a, Avenida Espa{\~n}a 1680, Valpara{\'i}so, Chile\\
$^{6}$Istituto Nazionale di Astrofisica, Osservatorio Astrofisico di Torino, Strada Osservatorio 20, I-10025 Pino Torinese, IT\\
}

\date{Accepted XXX. Received YYY; in original form ZZZ}

\pubyear{2022}

\begin{document}
%%%%%%%%%%%%%%%%%%%%%%%%%%%%%%%%%%%%%%%%%%%%%%%%%%%%%%%%%%%%
%\newpage
%\listoftodos %will add the list to the top of the pdf
%\newpage

%%%%%%%%%%%%%%%%%%%%%%%%%%%%%%%%%%%%%%%%%%%%%%%%%%%%%%%%%%%%
\label{firstpage}
\pagerange{\pageref{firstpage}--\pageref{lastpage}}
\maketitle

\begin{abstract}

The ability to automatically and robustly self-verify periodicity present in time-series astronomical data is becoming more important as data sets rapidly increase in size.
The age of large astronomical surveys has rendered manual inspection of time-series data less practical.
Previous efforts in generating a false alarm probability to verify the periodicity of stars have been aimed towards the analysis of a constructed periodogram.
However, these methods feature correlations with features that do not pertain to periodicity, such as light curve shape, slow trends and stochastic variability.
The common assumption that photometric errors are Gaussian and well determined is also a limitation of analytic methods.
We present a novel machine learning based technique which directly analyses the phase folded light curve for its false alarm probability.
We show that the results of this method are largely insensitive to the shape of the light curve, and we establish minimum values for the number of data points and the amplitude to noise ratio.

\end{abstract}


\section{Introduction}

The identification of periodic variable stars is not a trivial task;
well-understood statistical measures can be used to identify variability in time-series but not so easily periodic variability.
The Stetson variability index `\textit{I}'~\citep{Stetson1996} compares the variability of each observation with its neighbour and their errors.
The Von Neumann eta index `\textit{$\eta$}'~\citep{Neumann1941} represents the ratio of the mean of the successive differences squared, to the variance of the light curve.
Both of these methods are reasonably robust in detecting variability in time-series.
More simplistic methods, such as a comparison between some measure of scatter (Inter Quartile Range, Standard deviation $\sigma$ or Median Absolute Deviation) and the uncertainty, have also been shown to be useful~\citep{Sokolovsky2017}.
Using tools such as the Lomb-Scargle method~\citep{Lomb1976,Scargle1982} and Phase Dispersion Minimisation~\citep[PDM,][]{1978PDM}, we can construct a periodogram to probe for periodic variability.
Nevertheless, extrema in the periodogram are likely to be present regardless of whether or not the source is truly periodic. These extrema can scale with the amplitude of the periodic signal such that periodograms of periodic sources become distinct from truly random variability. However, in cases where a light curve features aperiodic or secular variability, ambiguities can arise \citep{Park_2021}. This is of particular issue when dealing with stars which can feature multiple sources of variability, such as asymptotic giant branch stars \citep{Templeton2005}, whose long term periodicity could be undifferentiable to that of secular variability by periodogram analysis alone.
Furthermore, their values do not scale universally (i.e.\ the peak value for an aperiodic source may be the same as that for a periodic source).

In cases where extrema are not present, this could be interpreted as an indication of insufficient periodogram coverage or the lack of periodic variability.

Thus, we do not automatically obtain a universal measure of periodicity from a periodogram.
If a periodogram shows candidate periods, then for smaller selections of sources, it is feasible to manually verify the periodicity of each. 
This is typically performed by visual inspection of the phase folded light curve.
Looking forward, in the current and future age of survey astronomy with surveys such as LSST~\citep{LSST2019}, ZTF~\citep{ZTF2019}, Kepler~\citep{Kepler2003} and TESS~\citep{TESS}, we anticipate time-series catalogues of sizes that render sufficient manual inspection increasingly non-viable.
Hence, a reliable and robust metric for identifying periodicity is required.

It is not a guarantee that a large survey will feature high cadence sampling.
Surveys such as VISTA Variables in the Via Lactea~\citep[VVV,][]{VVV, VVV2012}, the NEOWISE mission of the Wide Field Infrared Survey Explorer~\citep{WISE2010, Mainzer2014} and \textit{Gaia}~\citep{gaia_edr3_2} have catalogues which can also feature large sample sizes for which rigorous human inspection is impractical.
These surveys contain relatively few observations for each source, an issue that is also very common in small, targeted observing projects.
The sparse sampling makes it harder to confirm periodicity with classical methods.

The metric for determining periodicity in a time-series is commonly referred to as a False Alarm Probability (FAP).
Previous work on determining an accurate FAP has largely been directed toward the analysis of the constructed periodogram.
These methods, such as the method proposed by~\citet{Baluev2008}, employ extreme value statistics to determine an upper bound for the false alarm probability of a Lomb-Scargle periodogram.
This has the clear limitation that the method is designed to distinguish sinusoidal variations from Gaussian white noise, not accounting for stochastic variability, non-Gaussian errors, imprecise error estimates and non-sinusoidal periodic variations.
\citet{Baluev2009}~extended their earlier work to the case of multi-harmonic light curves but this is only a partial solution to the above issues.
Bootstrapping is another commonly used technique where the periodogram of a light curve that has been randomly shuffled N times to create N aperiodic periodograms is compared to that of the unshuffled light curve.
The FAP in this case is the percentage of times the peak of an aperiodic periodogram is larger than that of the peak from the suspect periodic periodogram.

In \citet{1978PDM}, a statistical analysis of the constructed PDM periodogram is used to obtain a metric of false alarm probability (P-value).
This method assumes that photometric errors are perfectly estimated Gaussians.
The absence of any other aperiodic variability is also assumed.
There is also no treatment of spurious artificial periodic signals, which can occur with unevenly and sparsely sampled light curves.
Many surveys feature these periods at varying rates of incidence. It is of particular note for ground-based surveys with semi-regular observing patterns, such as the VVV survey. 
Methods such as PDM that bin the phase-folded light curve to construct their periodogram are also limited by imperfections in the model.
This can become increasingly significant as sampling decreases.
This issue exists even with the binless approach to PDM presented by~\citet{plavchan2008}.
Separately, heuristic methods based on reduced $\chi^2$ statistics have been employed to distinguish true and false periodic variable star candidates~\citep[e.g.\ ][]{irwin2009}. This explicitly acknowledges the effects of an imperfect light curve model and imprecise photometric uncertainties.
In this work, we show how we can utilise neural networks to differentiate between true and false periodic variable star candidates without the need for a prior light curve model.

\section{Method}\label{sec:method}
In our approach, the analysis of the light curve is achieved via a Recurrent Neural Network~\citep[RNN,][]{lstm1997}.
An RNN was chosen because they are designed and used for serially correlated data, such as astronomical light curves.
Previous efforts in their use with light curves have shown their applicability and ability to parse astronomical time-series data~\citep{Burhanudin2021, zhang2018}. 
This network is trained on pre-labelled periodic and aperiodic phase-folded light curves of variable stars.
The network was trained for 96 epochs\footnote{An `epoch' here is an iteration over the whole training set} with an Adam optimiser~\citep{ADAM2014} and with 20\% of the training data used as a validation set.
Early stopping was used to halt training as soon as the incremental change in the validation loss function, $\Delta L < 10^{-5}$.

The model is constructed with 13 Gated Recurrent Unit~\citep[GRU,][]{cho2014} layers, 1024 nodes per layer and a binary cross entropy loss model.
The choice of GRUs over Long-Short Term Memory~\citep[LSTM,][]{lstm1997} was motivated by the calculated loss, which was lower for GRUs. The RNN was written in Keras~\citep{chollet2015keras}.

Ablative testing has shown that the specifics of the architecture of the network are not crucial and having `enough' GRUs is sufficient for operability.

\subsection{Data preparation}
The input training data consists of the magnitude ($m_i$), phase ($\phi_i$) and change in phase (i.e.\ $\Delta\phi_{i} = \phi_{i} - \phi_{i-1}$).
Magnitude errors were not used for this method as they commonly do not fully represent the true photometric uncertainty. 
We tested various combinations of features and removing the magnitude error consistently improved performance with lower loss and higher accuracy. 
Instead, we reject any points with a large magnitude error ($m_{i,err}\,\geq\,0.1$ in this case). We also reject points with a high DoPHOT \citep{DoPHOT1993} `Chi' parameter, which indicates a poor fit to a stellar profile.

The input also includes a feature that is derived from an interpolated fit of the time-series with 200 evenly spaced points, performed by an inverse distance-weighted K-nearest neighbours (KNN) regressor~\citep{knn} which was taken from Scikit-learn~\citep{sklearn}.
This was performed as a form of smoothing in an attempt to more clearly display variability with evenly spaced data. 

A randomly variable light curve will have an interpolated fit that tends towards a straight line. 
Each of these features were added after ablative testing (i.e.\ features were added and removed iteratively and the combination of features that produced the highest accuracy and lowest loss was used).
Each light curve was either cut to 200 data points in size or padded with zeroes to a length of 200.

The same light curve is phase shifted randomly 10 times by an amount between 0\,and\,$2\pi$ and each version is shown to the neural network.
This is done in an attempt to remove a dependency on the starting position of the light curve. 
This is similar to the methodology for contrastive learning~\citep{chen2020}.
We do not want the network to care about the absolute phase value.

Alternatively, we could ensure the light curve is always ordered from a set point in the light curve, such as the turning points.
However, we found this step to be unreliable with noisy data.
A single unfiltered outlier or otherwise erroneously extreme point would cause such an approach to fail as the light curve's minima could be incorrectly identified.

\section{Data}\label{section:trainingdata}
The training data used for training the neural network FAP (NN~FAP) is a combination of both real and synthetic light curves.

In the trained model used for this paper, there were 20\,000 real and 60\,000 synthetic light curves with half of each corresponding to periodic or aperiodic. 
This means that a FAP of 0 was given to the 10\,000 real and 30\,000 synthetic periodic light curves and a FAP of 1 was given to the other 10\,000 real and 30\,000 synthetic aperiodic light curves.
The synthetic light curves were split evenly across each of the five listed equations~(\ref{EQ:RR}--\ref{EQ:YSO}).

Through the development of this method, it was found that a small number of mislabelled light curves can have a large impact on the abilities of this method (i.e.\ an aperiodic light curve being labelled as periodic or vice versa).

\subsection{Real training data}

The training data are VVV light curves whose periodic nature was supported by classification from two optical surveys.
A set of 10,000 known real periodic light curves were identified by eye (by co-authors NM, CM \& WC) after cross-matching data from the VVV survey, (and a pre-release version of its time-series catalogue, VIRAC 2-$\beta$~\citep[][Smith et al., in prep]{VIRAC} with other known periodic variable star catalogues, namely the ZTF catalogue of periodic variable stars and the ASAS-SN catalogue of variable stars~\citep{ZTF2020, assasssn2019}.
The cross-matching was performed to generate a list of suspect periodic and aperiodic variable stars.

All of the 10,000 aperiodic light curves were identified by eye as rejected periodic variables.

\begin{figure}
	\includegraphics[width=\columnwidth]{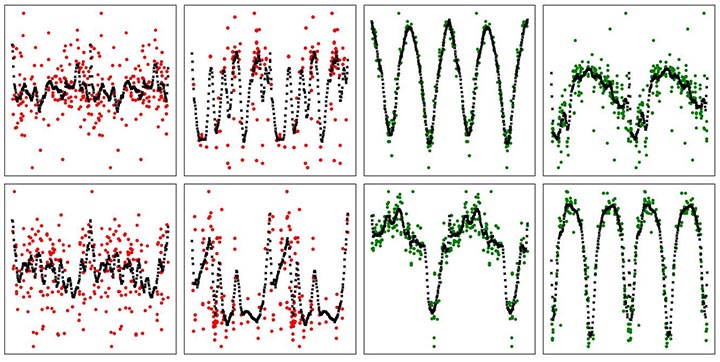}
    \caption{
        Some example real aperiodic (red) and periodic (green) light curves used for training. The black points represent the evenly spaced fit provided by the KNN regressor.
    }
    \label{fig:training_grid_example}
\end{figure}

Figure\,\ref{fig:training_grid_example} shows a random selection of real training light curves and their interpolated fit. Both the interpolated fit and the raw magnitude measurements are given to the RNN.

\begin{figure}
	\includegraphics[width=\columnwidth]{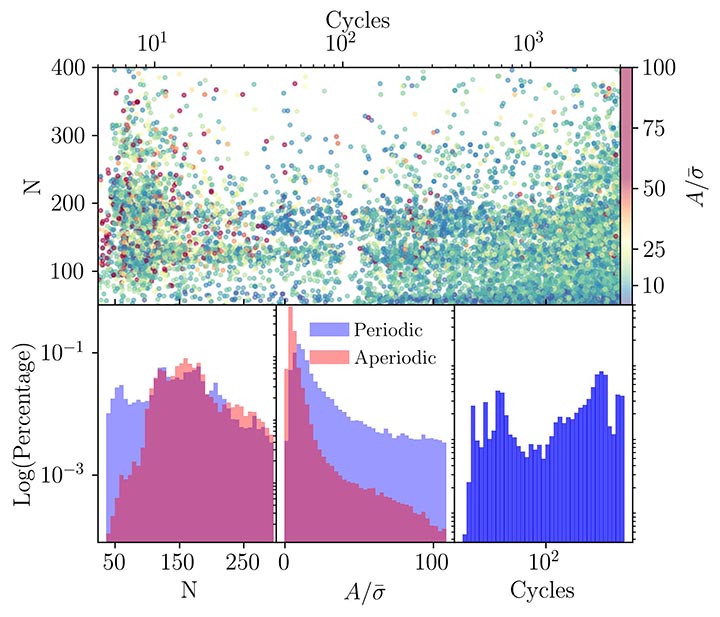}
    \caption{Showing the distribution of the number of data points (N), signal to noise ratio and cycles (light curve length / period) for the training set used. \textbf{Top:} Cycles versus the number of data points with colour axis as signal to noise ratio for periodic variables in the training data. \textbf{Bottom:} Histograms for each of these values.}
    \label{fig:training_data_hist}
\end{figure}

Figure\,\ref{fig:training_data_hist} shows the distributions of number of data points, signal to noise ratio and number of cycles in the time series for the training data. 
The real data is drawn directly from this distribution.
We show the number of cycles as opposed to period because the neural network is trained exclusively with phase folded light curves. As VVV light curves can vary in length, it is the product of the light curve length with the stars frequency that affects the structure of the phase folded light curve.

\subsection{Synthetic training data: Periodic light curves}

Synthetic light curves were created via the use of a real light curve with a periodic signal injected, similar to the work by~\citet{Graham2013}.  
An overview of the steps taken are as follows:

\begin{enumerate}
    \item Remove all photometric information from a real light curve, retaining only the time stamps.
    \item Inject periodic signal into `blank' light curve (see equations~\ref{EQ:RR}--\ref{EQ:YSO}).
    \item Generate the errors by sampling from those associated with the real photometry, using a look-up table.
    \item Scatter light curve based on the injected error.
\end{enumerate}

Training the neural network on exclusively sinusoidal light curves could bias our FAP against Eclipsing Binaries and other more complex light curves.

\begin{figure}
	\includegraphics[width=\columnwidth]{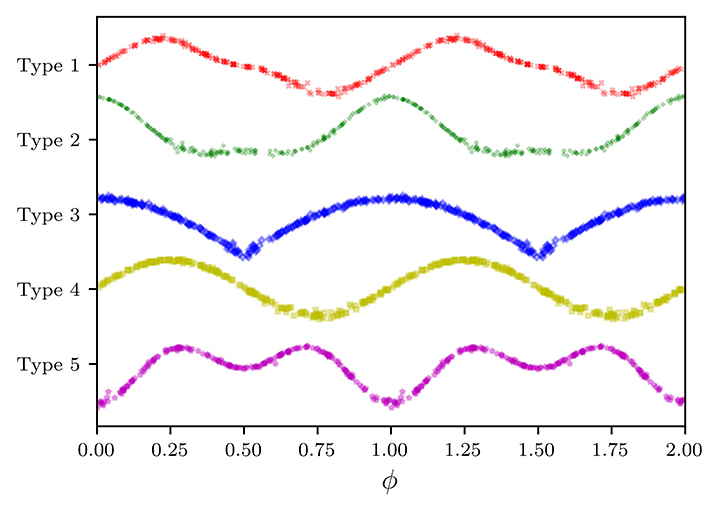}
    \caption{
        Examples of each of the forms of the light curves used for testing and training the neural network FAP.
    }
    \label{fig:class_examples}
\end{figure}

Figure~\ref{fig:class_examples} shows an example of each of the forms of light curves generated with 1000 measurements with an amplitude of 1\,mag. 
These equations aim to roughly (but not exactly or comprehensively) model the common types of pulsators and binary light curves that are seen~\citep{Molnar2022}. 
Type 1 is a distorted sinusoid which is a fairly standard form for synthetic light curves~\citep{Cincotta1995, Huijse2012}. Types 2\,\&\,5 are eclipsing binary-like light curves (i.e. more than one turning point per period). Type 3 is used to mimic the common identifying feature of a contact binary system~\citep{Kirk2016} and Type 4 is a simple sinusoid. An important reason for using multiple shapes to train the network is to remove as much of a dependency on light curve shape as possible.
This is similar to the methodology for contrastive learning. By showing the network multiple different shapes of a periodic signal we aim to remove any biases related to its shape.

The method by which the synthetic data is created also means that the light curve parameters are drawn from the distributions shown in Figure\,\ref{fig:training_data_hist}. 
The periods used are randomly selected from a uniform distribution between 0.1 and half the length of the light curve (period\,$\sim\!\mathbb{U}(0.1,\sim\!\!1500)$). 
We note that the period (or number of cycles) used for the synthetic light curves is largely inconsequential to how it is perceived by the RNN. 
The RNN is only shown the phase fold of the light curve and so there is little difference between otherwise identical light curves with different periods. This is also the case for the total time range of the light curve. Provided at least one cycle is captured, the number of measurements and signal-to-noise are the limiting factors. This is a potential caveat for this method as a low FAP could be assigned for a light curve with only one cycle, which is not sufficient for the actual identification of periodicity. We recommend only trusting the FAP from this method if the period is less then half of the length of the light curve (i.e. at least two cycles are captured).

\begin{equation}\label{EQ:RR}%this makes light curves that are like RR/Ceph
    \texttt{Type 1.}\,\,m(t) =  0.5sin \left ( \frac{2\pi t}{P} \right ) - B_{1} sin \left (\frac{4\pi t}{P} \right )- B_{2} sin \left (\frac{6\pi t}{P} \right )
\end{equation}

\begin{equation}\label{EQ:EB}%this makes light curves that are like EB/DN
    \texttt{Types  2 \& 5.}\,\,m(t) = 1 \pm \left (A_{1} sin \left ( \frac{2\pi t}{P} \right )^{2} + A_{2} sin \left (\frac{\pi t}{P}\right)^{2} \right )
\end{equation}

\begin{equation}\label{EQ:RECT}%full wave rectifies LC, some YSOs look like this?
    \texttt{Type 3.}\,\,m(t) =  \left |sin\left ( \frac{2\pi t}{P} \right ) \right |
\end{equation}

\begin{equation}\label{EQ:YSO}%sinusoidal, call it YSO because....
    \texttt{Type 4.}\,\,m(t) =  sin\left ( \frac{2\pi t}{P} \right )
\end{equation}

A periodic signal is added to the source light curve, and the photometric error is derived using a KNN search of a dataset containing information about the photometric uncertainty of 1\,000\,000 data points from the VIRAC database. This dataset is utilised to identify the 100 nearest neighbours, from which the mean and standard deviation are computed.

Each data point in the light curve has its photometry ($m$) and photometric error ($m_{err}$) drawn from a Gaussian constructed of these 100 nearest neighbours.

\subsection{Synthetic training data: Aperiodic light curves}

We employ two methods to generate aperiodic light curves:
a real or synthetic periodic variable has its photometric order randomly shuffled.
The time data is left unmodified to conserve the observing cadence of the original survey.
We effectively create a light curve of random noise with the survey's observing pattern conserved.
This method also removes any other correlated effects, such as photometric uncertainty, that may be present in real aperiodic light curves.
One caveat present is that by destroying correlated effects, the neural network could differentiate between the aperiodic and periodic synthetic light curves with greater ease.
The second method of aperiodic synthetic light curve generation involves taking a known non-variable star (identified with a Stetson index ${<}0.1$) and re-sampling the photometric points with a larger scatter.
For each measurement a Gaussian is constructed with $\mu=m_i$ and $\sigma \geq 3\,\times\,m_{i,err}$, where $m_{i,err}$ is the measurement error.
The light curve is then re-scaled to ensure a realistic amplitude.

This method retains as much temporally correlated, but non-periodic, information as possible compared to the random shuffle method. 
An example of this is with astronomical seeing, which can vary on long timescales, affecting multiple measurements. With VVV (and subsequent catalogue VIRAC 2-$\beta$) data we have instances where bad seeing causes DoPHOT to systematically underestimate flux in crowded fields. Such a case could appear as a non-periodic signal in the light curve. 
In less crowded fields, poor weather will increase the uncertainty at times, creating correlated uncertainty which may occasionally lead to a spurious aperiodic signal. This is of particular note as the neural network is never shown the photometric uncertainty. 
This method of inflating measurement error will weaken but not fully destroy these correlated effects.

The random shuffle method enables training with non-Gaussian aperiodic signals. 
Due to the limitations of these methods, it is beneficial to also have real training data. The synthetic data has the advantage of volume with the certainty of aperiodicity. This allows us to construct a training data set large enough to train an RNN. 

\subsection{Test Data}\label{sect:testdata}
\begin{figure}
	\includegraphics[width=\columnwidth]{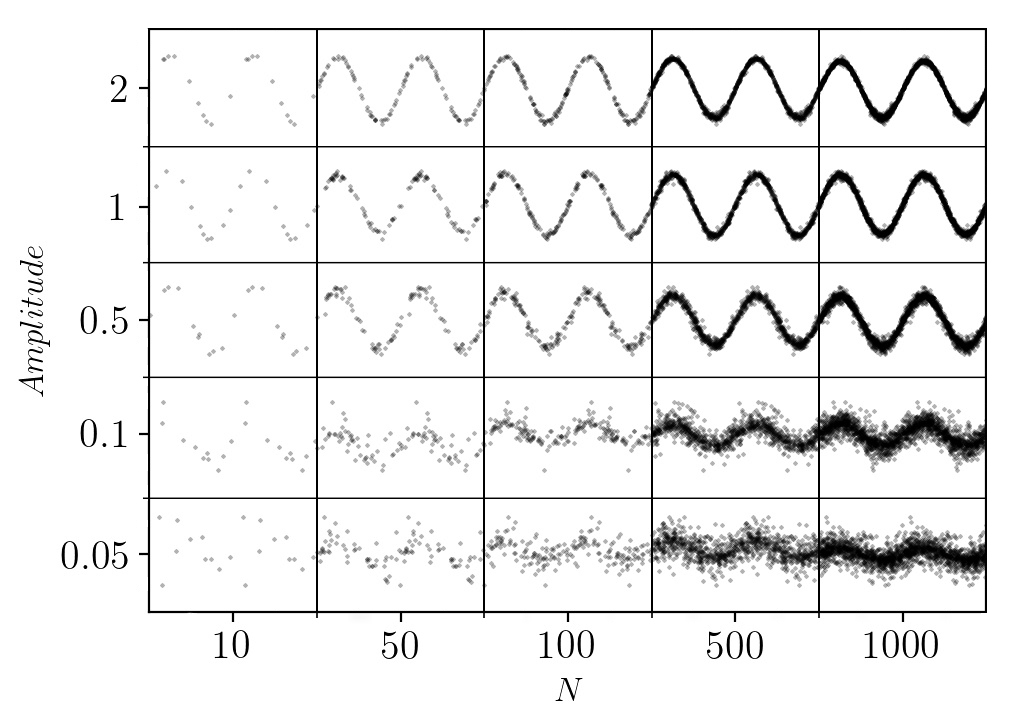}
    \caption{
        Examples of a synthetic sinusoidal light curve varying through the number of data points in the light curve on the x-axis and the amplitude of the light curve in the y-axis.
        The median magnitude error for each point was 0.1.
    }
    \label{fig:N_vs_Amplitude}
\end{figure}

We generate 3 data sets to test our classifier.
A real data set was constructed by manually classifying 8000 previously unseen real light curves taken from the same VVV survey. 
These 8000 sources were identified from the same ZTF and ASAS-SN periodic catalogues that were used in training.
Each light curve has a $A/\bar{\sigma} > 2$ (where `$A$' is the amplitude calculated as the difference between the 1\% and 99\% percentile after sigma clipping and `$\bar{\sigma}$' is calculated as the median value of the magnitude error.) 
The manual classification of the real light curves involved selecting phase-folded periodic variables by eye.
This was independently repeated multiple times by three astronomers to ensure reliability.
All of the astronomers agreed on classification.
Any ambiguous light curves were removed from the set.
Two synthetic data sets were also constructed via the method described in Section\,\ref{section:trainingdata}. 
The data set `Variable N' was generated as 80\,000 identical synthetic light curves with only the number of data points per light curve varied ($10<\,\mathrm{N}\,<600$).
A median SNR ($A/\bar{\sigma}$) of 10 was generated for each of these. 
The data set `Variable SNR' was generated as 80\,000 identical synthetic light curves with only the signal-to-noise ratio varied.
For each light curve in the Variable SNR data set, there were 200 data points used. 
Figure~\ref{fig:N_vs_Amplitude} exemplifies both `Variable N' and `Variable SNR' on the $x$ and $y$-axes, respectively.
The four types of synthetic variables used were evenly split for both of the synthetic data sets.

\section{Experimental results from RNN}

To quantify the performance of the NN~FAP we can test its ability as a binary classifier and compare it to the commonly used Baluev method.
We use the generalised Lomb-Scargle periodogram along with its associated FAP as described by~\citet{GLS2009} for our calculations of the Baluev~FAP.
The Baluev~FAP typically lies in a range between unity and $10^{-200}$ and so the y-axis of the Baluev FAP plots have been shown as both linear and logarithmic scaling.
%To more clearly compare the two methods, a normalisation of $\mathrm{FAP}_{\mathrm{Bal}} = (log_{10}(\mathrm{FAP}_{\mathrm{Bal}}) + 200) / 200$ is applied to all plots showing the Baluev~FAP.

\subsection{Performance Measurements}
The Receiver Operator Characteristic (ROC) curve is used to measure the capability of a binary classifier as the threshold for classification is varied.
An idealised binary classifier will have a threshold at which the \textit{sensitivity} and \textit{specificity} are equal to 1.

\begin{figure}
	\includegraphics[width=\columnwidth]{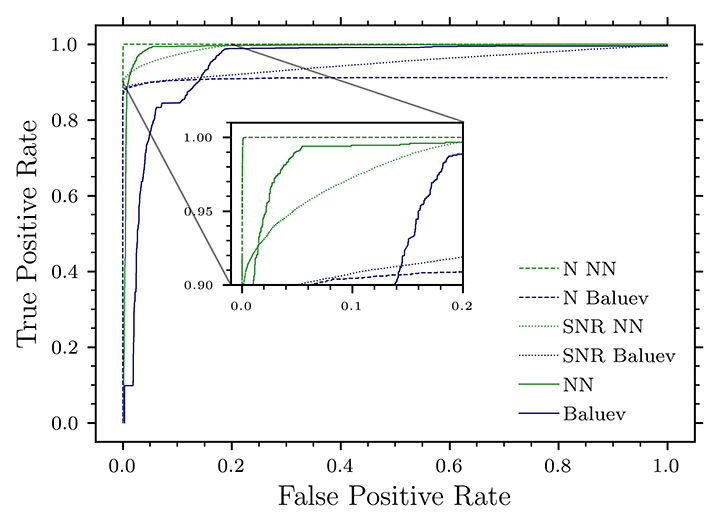}
    \caption{
        Showing the ROC Curve for the neural network and Baluev methods as a binary classifier. 
        \textbf{Solid line} real data set classified by eye.
        \textbf{Dashed line} synthetic data set where the number of measurements was varied (Figure~\ref{fig:fap_vs_n}).
        \textbf{Dotted line} a synthetic data set where the SNR was varied (Figure~\ref{fig:fap_vs_a}).
    }
    \label{fig:AUC}
\end{figure}

Figure~\ref{fig:AUC} shows the true positive rate (otherwise known as the sensitivity) versus the false positive rate (otherwise known as 1\,-\,specificity).
Equation~(\ref{EQ:sens}) shows more clearly how sensitivity and specificity are defined (where TP and TN are True Positive and Negative respectively.
FP and FN are False Positive and Negative respectively.) 
\begin{equation}\label{EQ:sens}
    \mathrm{Sensitivity} = \frac{TP}{TP + FN}
    \,\,\,\mathrm{Specificity} = \frac{TN}{TN + FP}
\end{equation}

 The Area Under the ROC Curve (AUC) can be calculated as an evaluation metric for a binary classifier.
 Table~\ref{tab:AUC} shows the AUC for each tested data set.
 This shows that the NN~FAP method has a larger AUC for each data set than the Baluev method.
 This indicates that the NN~FAP method performs better in each test.
 However, the AUC metric does not tell the whole story, as discussed below.
 
 \begin{table}
     \centering
     \begin{tabular}{l|c|c}
     \hline
        Source & NN~FAP & Baluev\\
        \hline \hline
        Real & 0.99193 & 0.95245\\
        Variable SNR & 0.99808 & 0.97843\\
        Variable N & 0.99703 & 0.97393\\
        \hline\\
     \end{tabular}
     \caption{
        Showing the AUC for each data set and method.
     }
     \label{tab:AUC}
\end{table}

\begin{figure}
	\includegraphics[width=\columnwidth]{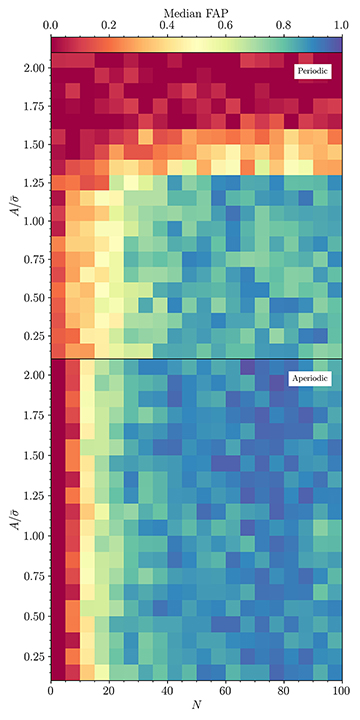}
    \caption{Showing the FAP calculated for synthetic light curves as a function of the number of data points `N' in the x-axis and $A/\bar{\sigma}$ in the y-axis}
    \label{fig:NA_faps}
\end{figure}

Figure\,\ref{fig:NA_faps} shows the median NN FAP as a function of N and $A/\bar{\sigma}$. This was calculated for 80,000 synthetic sinusoidal periodic light curves (i.e. generated with equation\,\ref{EQ:YSO}).
The calculations were performed with a range of $3\,<\,\mathrm{N}\,<\,100$ and $0.1\,<\,A/\bar{\sigma}\,<\,2.1$. We can see that this method appears reliable provided $A/\bar{\sigma}\,>,1.5$ and $\mathrm{N}\,>40$. 
We note that a median value does not reveal occasional failures and we suggest a limit of $\mathrm{N}\,>50$ for greater reliability, based on the results in section\,\ref{sec:fapvn} and Figure\,\ref{fig:fap_vs_n}.
A small amplitude with respect to the uncertainty is likely to give false negatives whereas a small number of measurements is likely to give false positives.

We also randomly selected 1000 eclipsing binary stars from the VIVACE catalogue~\citep{Molnar2022}. This catalogue was generated from the same VVV data that this model was trained on. All light curves were independently verified as eclipsing binary for this test. We construct a periodogram with both Lomb-Scargle and PDM and choose whichever period produced the lowest NN FAP. We find that 997 of the 1000 were identified as periodic with a FAP~<~0.1. The three light curves which failed to be identified each featured a FAP~>~0.6. In each of these three light curves there was a significantly shorter transit time paired with N~<~60 measurements. The identification of the correct periodicity can be an issue when a light curve can look periodic when phase folded at multiple different periods. If an eclipsing binary features a similar size and shape for each eclipse then the NN FAP can erroneously be assigned half the true period as the two dips in the light curve are likely to be undifferentiable in the phase fold. This can be problematic for equal mass eclipsing binary systems.

\subsection{FAP vs N}\label{sec:fapvn}
\begin{figure}
	\includegraphics[width=\columnwidth]{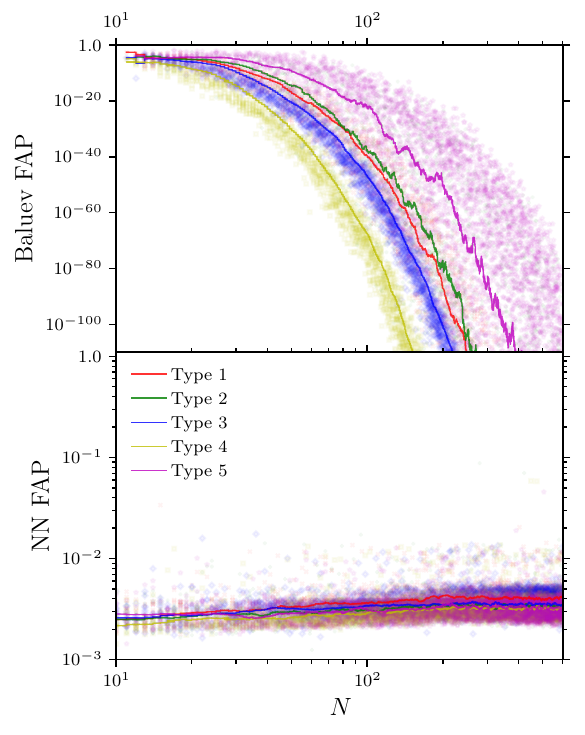}
    \caption{
        Both false alarm probabilities versus the number of measurements in the synthetic periodic light curve described in section\,\ref{sect:testdata}
        The colours and markers correlate to those shown in Figure~\ref{fig:class_examples}. \textbf{Top}\,: Baluev~FAP versus $N$.
        \textbf{Bottom}\,: NN~FAP versus $N$.
    }
    \label{fig:faps_vs_n}
\end{figure}

The number of measurements used to constitute a light curve can vary by orders of magnitude dependent on the survey.
Surveys such as Kepler and TESS feature highly sampled light curves which should not pose an issue to any FAP technique.
However, this is not always the case and many surveys feature light curves with fewer than 100 measurements.
Figure~\ref{fig:faps_vs_n} shows how the Baluev~FAP and the NN~FAP vary as a function of the number of measurements `N' for the synthetic light curve described in section\,\ref{sect:testdata}.
The NN~FAP does not produce any significant number of false negatives as the number of measurements decreases to 10.
The Baluev~FAP has a clear trend as a function of N and starts to increase to a problematic range of values as N approaches ${\sim}$50 measurements.
It can also be seen that the Baluev~FAP has a dependency on the shape of the light curve with more sinusoidal light curves assigned a lower FAP compared to more complex light curves such as eclipsing binary shapes.
This is an issue as it can lead to incorrect conclusions on the demographics of variable stars. 

\begin{figure}[h]
     \centering
    \begin{subfigure}[b]{\linewidth}
         \includegraphics[width=\columnwidth]{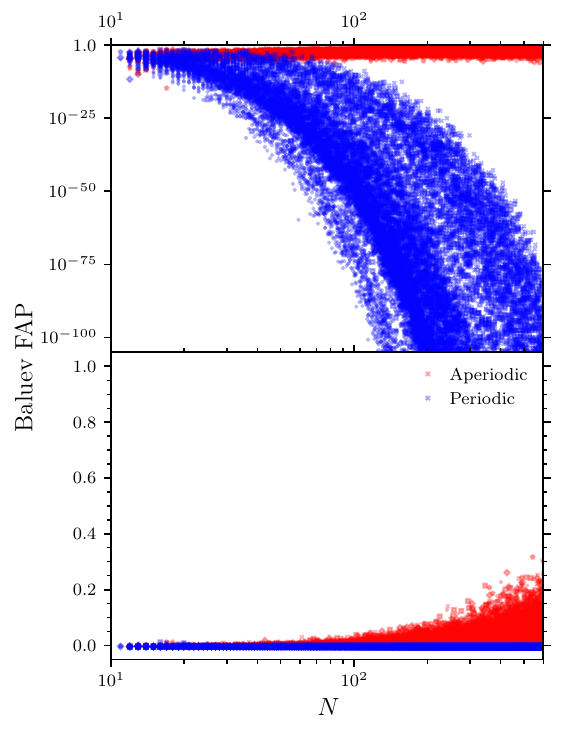}
         \label{fig:bal_vs_n}
     \end{subfigure}
    \begin{subfigure}[b]{\linewidth}
         \centering
         \includegraphics[width=\columnwidth]{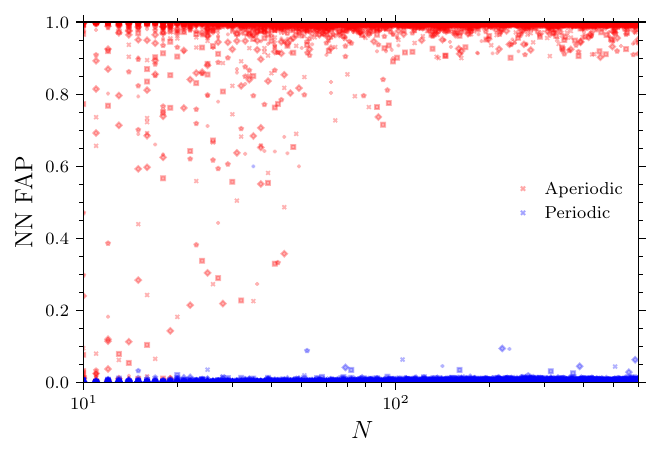}
         \label{fig:nn_vs_n}
     \end{subfigure}
        \caption{
            Both false alarm probabilities versus the number of measurements in the synthetic light curves described in section\,\ref{sect:testdata}.
            The red points show the FAP assigned to aperiodic light curves and the blue shows periodic light curves.
            It can be seen that the Baluev~FAP is more likely to assign false negatives whereas the NN~FAP is more likely to assign false positives. The Baluev FAP rarely exceeds 0.1 and never approaches unity.
            \textbf{Top}\,: Baluev~FAP versus $N$.
            \textbf{Bottom}\,: NN~FAP versus $N$.
            Each light curve here featured $A/\bar{\sigma} = 10$.
            The marker shapes correspond to those shown in Figure~\ref{fig:class_examples} (i.e. a cross represents `Type 1' and a plus `Type 2'...).
        }
        \label{fig:fap_vs_n}
\end{figure}

Again using the synthetic light curves described in section\,\ref{sect:testdata}, in Figure~\ref{fig:fap_vs_n} (bottom panel) we show that the NN~FAP sometimes falls to low values for aperiodic light curves as $N < 50$, potentially leading to false positive classifications. These false positives arise as any variable light curve with a small number of points will more easily represent a periodic light curve at a given phase fold. Caution should be taken with this method when searching for periodic variables with fewer than 50 measurements.
By contrast, the Baluev~FAP does not suffer from this problem but Figure~\ref{fig:fap_vs_n} (top panel) shows that it is more likely to assign false negatives to periodic light curves within the same range.

It is not possible to define a threshold for either method which we can use to perfectly separate the periodic and aperiodic light curves. Such a threshold must be set by the user depending on preference regarding completeness and purity.
The periodic light curves shown in Figure\,\ref{fig:fap_vs_n} have a maximum NN FAP of 0.791 but the minimum NN FAP for aperiodic light curves is 0.01.
The Baluev FAP has a maximum value of 0.015 for periodic light curves but a minimum value of \(1.197\times10^{-15}\) for the aperiodic light curves. The Baluev FAP values for periodic and aperiodic light curves overlap despite never approaching 1.
The median Baluev FAP for the aperiodic light curves when \(N\leq100\) is 0.0012 and when \(N\leq50\) it is 0.0005. 
Using the widely adopted criterion for the Baluev FAP of \(log_{10}(FAP)<-2\)\,\citep{Koeltzsch2009,Herbst2000,chen2020,Botan2021} yields misidentification of only four of the synthetic aperiodic light curves plotted in Figure\,\ref{fig:fap_vs_n} as periodic, while incorrectly categorising 13,849 (\(46.4\%\)) aperiodic stars as periodic. This indicates that a lower threshold is more suitable for our synthetic light curves.
In \citet{Molnar2022} a Baluev FAP selection of \(log_{10}(FAP)<-10\) was used to define a reliable but incomplete set of VVV light curves for training.
If we were to use that cut for this data we would misidentify 1152 (\(3.86\%\)) periodic light curves as aperiodic and 502 (\(1.68\%\)) aperiodic light curves as periodic.
If we use a NN FAP of 0.15 we misidentify 1 periodic light curve as aperiodic and 400 (\(1.34\%\))  aperiodic light curves as periodic.

False positives will arise, or not, depending on the FAP threshold value that is adopted.
The NN~FAP method performs very well in the AUC test for Variable N (see Table~\ref{tab:AUC}) because, even where aperiodic light curves have a low FAP, the periodic light curves have even lower FAP values.
This allows the binary classifier to be successful, in principle, if the threshold FAP could be ideally selected.
However, in practice, this will rarely be possible.

\subsection{FAP vs Amplitude}

The signal-to-noise of a light curve is a common source of erroneous periodicity classification.
Periodic variable stars can host a range of amplitudes depending on the source of variability.
As such, it is not uncommon to investigate variable stars whose variability is similar to, or below, the photometric uncertainty. 

\begin{figure}
	\includegraphics[width=\columnwidth]{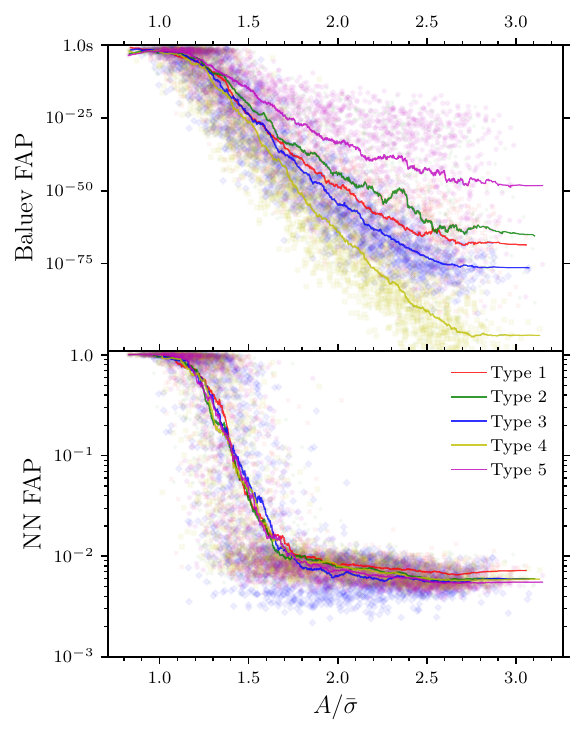}
    \caption{
        Both false alarm probabilities versus the amplitude of a synthetic periodic light curve divided by the median average of the photometric uncertainty.
        The colours and markers are the same as that shown in Figure~\ref{fig:class_examples}.
        \textbf{Top}\,: Baluev~FAP versus $A/\bar{\sigma}$.
        \textbf{Bottom}\,: NN~FAP versus $A/\bar{\sigma}$.
    }
    \label{fig:fap_vs_a}
\end{figure}

It can be seen in Figure~\ref{fig:fap_vs_a} that both the NN and Baluev~FAP feature a dependency on $A/\bar{\sigma}$.
Both the NN~FAP and the Baluev~FAP suffer from false negative rates as $A/\bar{\sigma} \rightarrow 1.5$.
Again, it can be seen that the NN~FAP does not suffer from a structure-dependent FAP, unlike the Baluev~FAP.
This is not surprising as the Lomb-Scargle method, to which the Baluev~FAP is applied, is effectively a sinusoidal fitting method and hence will feature such structure-based dependencies. 

The median Baluev FAP for the periodic light curves when \(A/\bar{\sigma}\leq1.25\) is 0.011 and 0.020 for aperiodic sources. The NN FAP at the same \(A/\bar{\sigma}\) has a median value of 0.959 for periodic sources and 0.998 for aperiodic sources. 
Both methods feature a significant level of confusion at such a low \(A/\bar{\sigma}\) but they do so at different absolute values, the Baluev FAP rarely features values larger than 0.1.
If we use the same value of \(log_{10}(FAP)<-10\) from \citet{Molnar2022} for the Baluev FAP we misidentify 2028 (\(3.86\%\)) periodic light curves as aperiodic. 
If we use a NN FAP of 0.15 we misidentify 1996 (\(3.79\%\)) periodic light curve as aperiodic. Neither method misidentifies any aperiodic sources as periodic sources with this selection.

\begin{figure}
     \centering
    \begin{subfigure}[b]{\linewidth}
         \centering
         \includegraphics[width=\columnwidth]{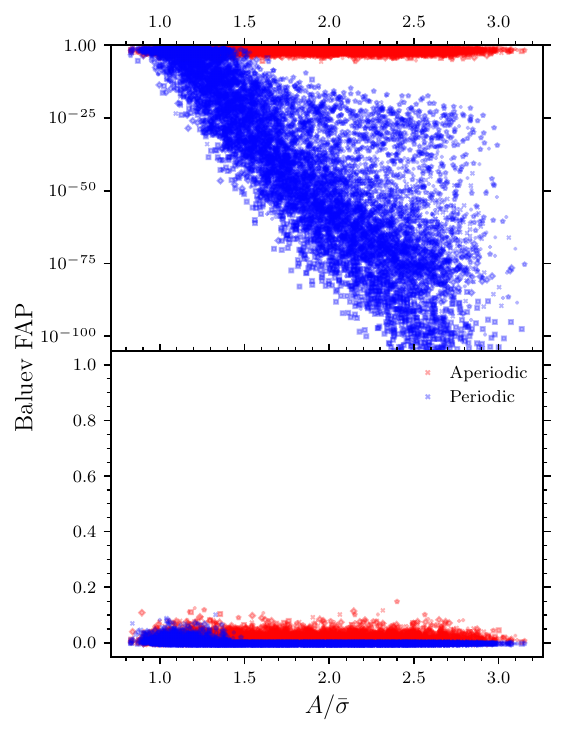}
         \label{fig:bal_vs_a}
     \end{subfigure}
    \begin{subfigure}[b]{\linewidth}
         \centering
         \includegraphics[width=\columnwidth]{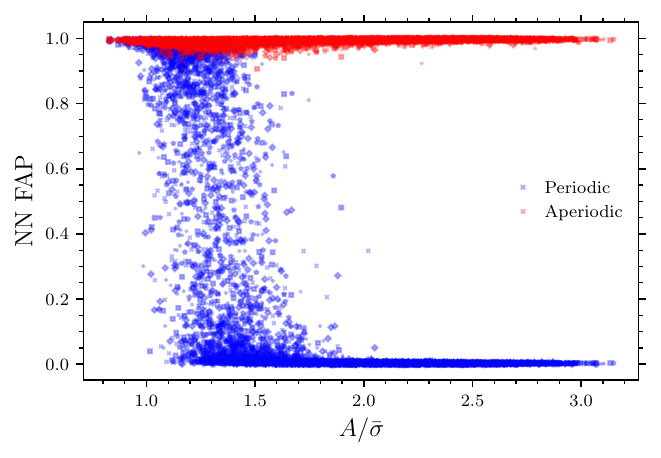}
         \label{fig:nn_vs_a}
     \end{subfigure}
        \caption{
            Both false alarm probabilities versus the amplitude of synthetic periodic and aperiodic light curves divided by their median average of the photometric uncertainty.
            Both methods show how their reliability begins to fail at $A/\bar{\sigma} \approx 1.5$.
            \textbf{Top}\,: Baluev~FAP versus $A/\bar{\sigma}$.
            \textbf{Bottom}\,: NN~FAP versus $A/\bar{\sigma}$.
            200 data points were used for these light curves.
            The marker shapes correspond to those shown in Figure~\ref{fig:class_examples}.
        }
        \label{fig:fap_vs_aa}
\end{figure}

\section{Testing with other surveys}

\begin{figure}
	\includegraphics[width=\columnwidth]{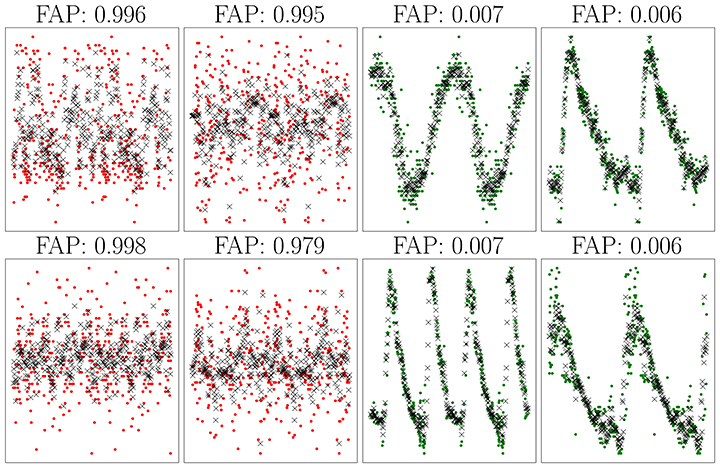}
    \caption{Randomly selected examples of identified periodic (green, right) and aperiodic (red, left) variable stars found in the CRTS survey. Each subplot displays the assigned NN FAP as its title. The green and red points represent the raw magnitude as a function of phase for the periodic and aperiodic light curves, respectively. The black points represent the KNN interpolated fit to the raw light curve. The Baluev FAP for each of the aperiodic sources was above $2\times10^{-5}$ and the periodic sources were all below $1\times10^{-60}$.}
    \label{fig:crts_periodics}
\end{figure}

\begin{figure}
	\includegraphics[width=\columnwidth]{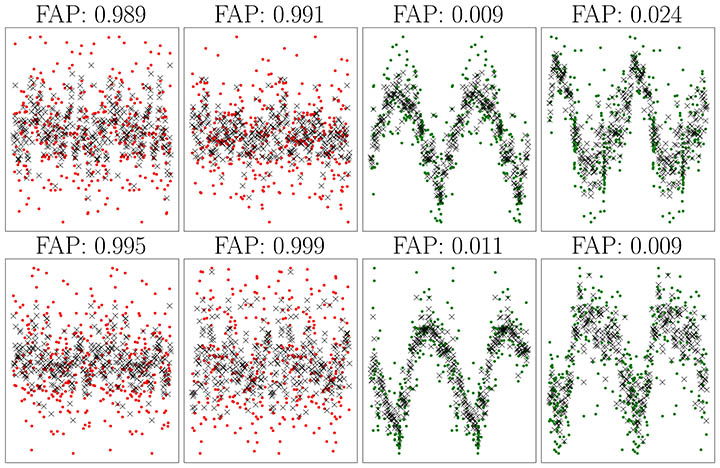}
    \caption{A random sample of identified periodic (green) and aperiodic (red) variable stars found in the ZTF survey. With each subplot showing the assigned NN FAP as its title. The Baluev FAP for each of the aperiodic sources was above $2\times10^{-14}$ and the periodic sources were all below $1\times10^{-51}$.}
    \label{fig:ztf_periodics}
\end{figure}

\begin{figure}
	\includegraphics[width=\columnwidth]{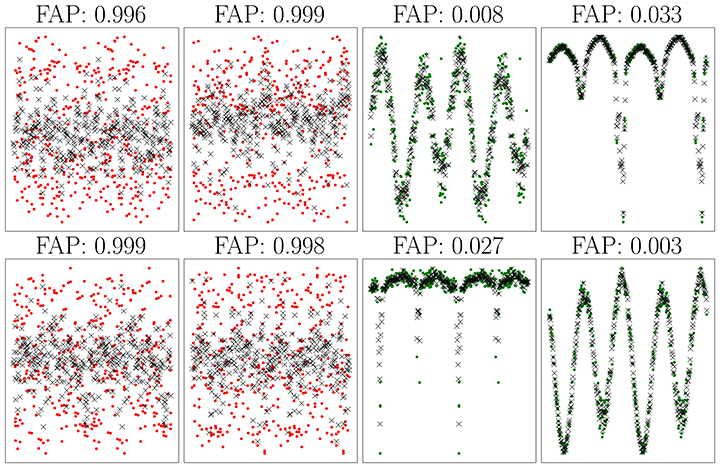}
    \caption{A random sample of identified periodic (green) and aperiodic (red) variable stars found in the Kepler survey. Each subplot shows the assigned NN FAP as its title. The Baluev FAP for each of the aperiodic sources was above $2\times10^{-4}$. The bottom left periodic variable has a Baluev FAP of $2.746\times10^{-6}$. The other periodic sources were all below $1\times10^{-43}$ but each had an incorrect period of half the true period.}
    \label{fig:kepler_periodics}
\end{figure}

\begin{figure}
	\includegraphics[width=\columnwidth]{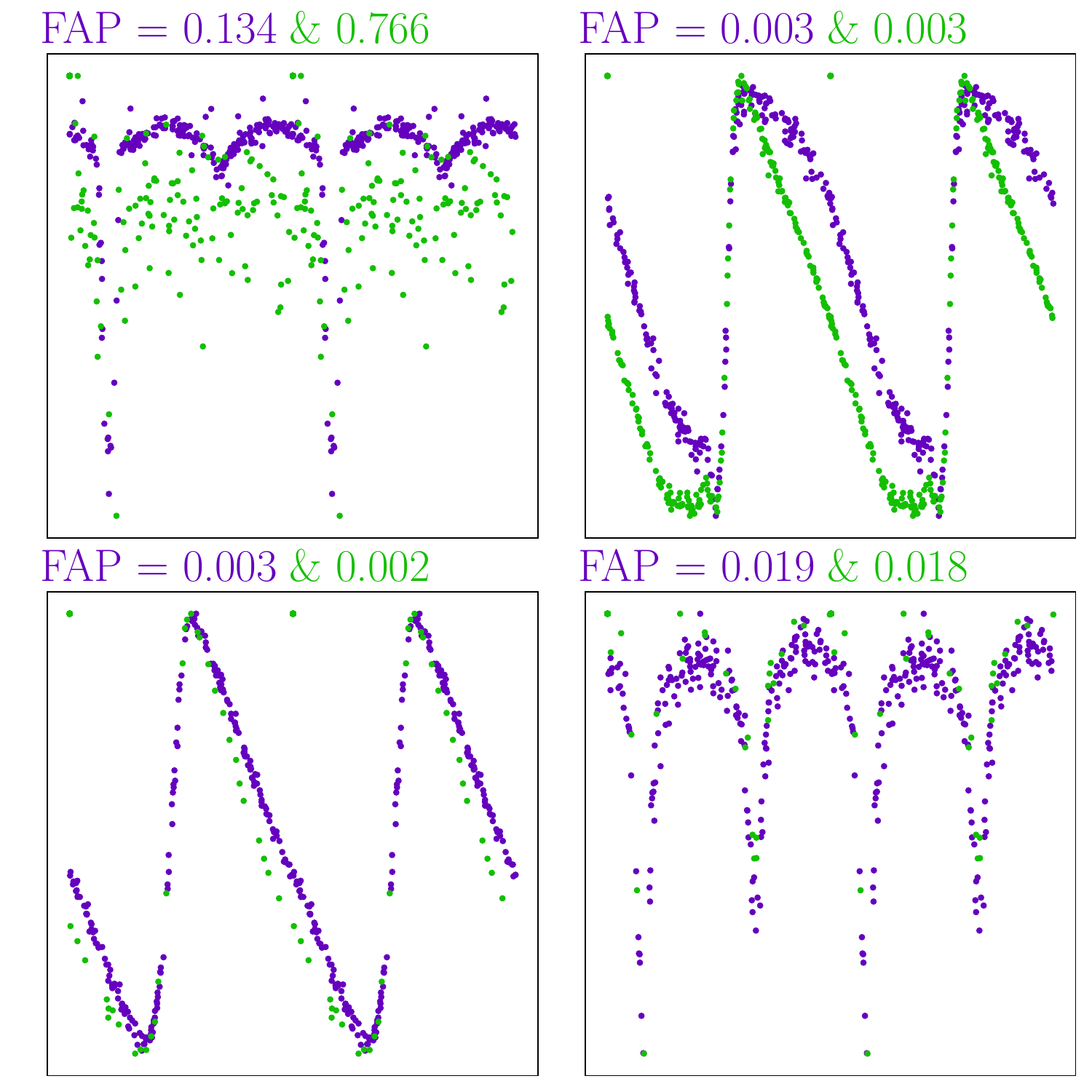}
    \caption{Periodic variables from the OGLE selection of variable stars. The green points represent the light curve of the star in the `V' filter and the purple represent the `I' filter. Three of the stars are identified as periodic in both V and I filters. The top left panel (OGLE-BLG-ECL-124368) was not identified as clearly periodic at any period in `V' and a higher NN FAP was given (although below that found for the aperiodics in~\cref{fig:crts_periodics,fig:ztf_periodics,fig:kepler_periodics})
    From top left to bottom right the Baluev FAPS are $0.967$ for `V' and $0.9$ for the incorrect period in `I', $9\times10^{-60}$ in `V' and $1\times10^{-235}$ in `I', $0.016$ for the incorrect period in `V' and $9.51\times10^{-141}$ in `I', $0.160$ in `V' and $6.9\times10^{-65}$ for the incorrect period in `I'}
    \label{fig:ogle_periodics}
\end{figure}

Our proposed method of calculating a FAP is universal and independent of the method of period detection. We can also show that the NN FAP method can be applied to data that is not drawn from the same distribution as the training data. 
Figure\,\ref{fig:crts_periodics} shows periodic and aperiodic variable stars in the CRTS \citep{crts}. Figure\,\ref{fig:ztf_periodics} shows them for ZTF. Figure\,\ref{fig:kepler_periodics} shows them for kepler. Figure\,\ref{fig:ogle_periodics} shows periodic variables in OGLE\,\citep{ogle2015} data. Each of the sub-plots in these figures show the assigned NN FAP. 

The top left panel of figure\,\ref{fig:ogle_periodics} shows the light curves in two filters for the source "OGLE-BLG-ECL-124368" which appears much more clearly periodic in `I' than `V'. The NN FAP reflects this, showing a higher FAP for the `V' band data. The `V' band data does still show a poorly sampled transit at the phase folded period, hence the NN FAP is above 0.5 but below 0.9.
The model used to identify these variables was trained as described in section\,\ref{section:trainingdata} with VVV light curves. The periodicity of each of these stars was identified by choosing the period which produced the lowest NN FAP extracted from a PDM periodogram. 
For both the CRTS and ZTF light curves the Baluev FAP was sufficient for differentiating between aperiodic and periodic variable stars. For three of the Kepler light curves the Lomb-Scargle periodogram incorrectly assigned half of the period with the a low Baluev FAP.
One of the Kepler light curves was not identified as periodic by the Baluev FAP (Bottom left panel of periodic variables in figure\,\ref{fig:kepler_periodics}). Only one of the OGLE light curves was correctly identified as periodic in both `V' and `I' by the Baluev FAP (Top right panel in figure\,\ref{fig:ogle_periodics}) although with a notably different FAP of $9\times10^{-60}$ in `V' and $1\times10^{-235}$ in `I'.
Each of the other OGLE light curves were either incorrectly given half of the true period or given a Baluev FAP indicative of aperiodic variability. The Lomb-Scargle periodogram also correctly identified the `I' band period of the bottom left panel with a Baluev FAP of $9.51\times10^{-141}$ but failed to extract the correct period for `V' band. The Lomb-Scargle periodogram and Baluev FAP predominantly struggled with more complex eclipsing binary shaped light curves.

\section{FAP Periodogram}\label{section:periodograms}

\begin{figure}
    \centering
    \includegraphics[width=\linewidth]{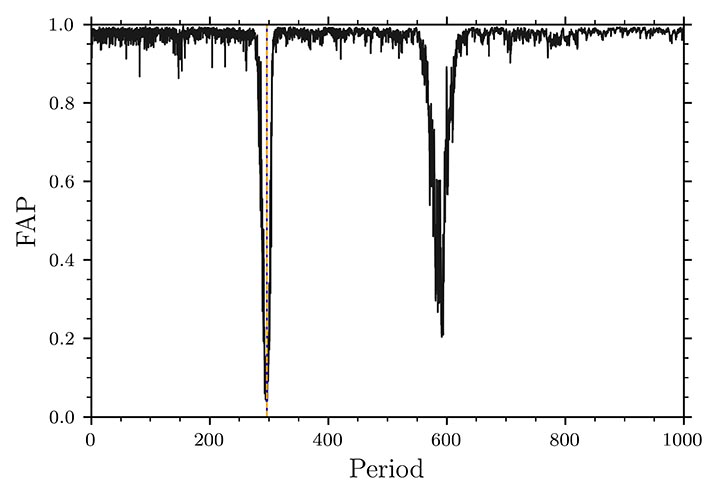}
    \caption{
        A periodogram constructed from the NN~FAP method.
        The periodogram took 23 minutes on 64 cores to compute.
        The correct minima is identified despite the binary-like construction of the periodogram.
    }
    \label{fig:fappsd}
\end{figure}

The NN FAP method presented above can be seen as something analogous to a neural network version of the PDM method so we can try to use it as such, i.e.\ for the construction of a periodogram rather than false alarm probability calculation.
We can calculate a FAP for a set of trial periods and the period which returns the lowest FAP should be the correct period.
This has the added benefit of generating a periodogram on a universal scale and thus the FAP is given along with the periodogram.
Currently, this approach is limited by its computationally demanding nature.
Future developments in computing paired with this method being modified for periodogram construction purposes will make this work more practical.
Figure~\ref{fig:fappsd} shows the periodogram constructed for a synthetic light curve (of type 5, Eq.\,\ref{EQ:EB}) with 200 points, a SNR of 2 and a period of 296.4 days. 
This periodogram took 23 minutes to construct and correctly extracted the correct period (inference was run on 64 CPU cores).
This compares to the 0.2 seconds it took for the PDM method to construct the same periodogram and achieve the same results (without a FAP). 
Both the NN periodogram and the PDM periodogram suffered from aliasing at multiples of the true period but both also correctly assigned the true period the largest peak value.

\section{Conclusions}
We have shown that utilising the flexibility afforded by neural networks allows a more robust analysis of light curves.
Using synthetic and real data, RNNs can be trained to produce a reliable and universal measure of periodicity. 
A study of the parameter space (namely the signal-to-noise ratio and temporal density) demonstrated how and when this method fails in comparison with the commonly used Baluev method.
This method remains reliable where $N > 50$ with $A/\Bar{\sigma} > 10$ or $A/\Bar{\sigma} > 1.5$ with $N \geq 200$. 
As we analyse the phase-folded light curve and not the periodogram, the NN~FAP is independent of the tools used to construct the periodogram.
This method is more analogous to a universally scaled PDM and so the network is effectively analysing the structure of the phase-folded light curve. This has further implications for a possible method of period detection that were explored in Section~\ref{section:periodograms}.

Figure~\ref{fig:AUC} and Table~\ref{tab:AUC} have shown how this method outperforms the Baluev method for both synthetic and real data.
Given a data set for candidate periodic variable stars, this method will provide a more complete search for periodicity, at the expense of occasionally generating more false positives for small N.

We highlight that the most challenging aspect of this method is the data preparation which is outlined in Section~\ref{section:trainingdata}.
Care must be given to how the training data is constructed and prepared.
This method is provided both with the ability to retrain on different data sets as well as pre-trained with the data described above.
We expect the method to be fully functional in its pre-trained state within the parameters outlined in this paper.
Conversely, this is not the case for the network's architecture which was shown by ablative testing to be relatively inconsequential to the performance.

\section*{Acknowledgements}

NM and WJC are funded by University of Hertfordshire studentships; furthermore NM recognises the computing infrastructure provided via STFC grant ST/R000905/1 at the University of Hertfordshire. CM acknowledges support from the UK's Science and Technology Facilities Council (ST/S505419/1).
ZG is supported by the ANID FONDECYT Postdoctoral program No. 3220029. ZG and AB acknowledge support by ANID, -- Millennium Science Initiative Program -- NCN19\_171. The authors would like to thank Mike Kuhn for his help in proving of ZTF light curves and helpful comments. 

%%%%%%%%%%%%%%%%%%%%%%%%%%%%%%%%%%%%%%%%%%%%%%%%%%
\section*{Data Availability}
The code used in this project is available at~\href{https://github.com/nialljmiller/NN_FAP}{github.com/nialljmiller/NN\_FAP}.
The model and weights are hosted at ~\href{www.kaggle.com/data sets/niallmiller/nn-fap}{www.kaggle.com/data sets/niallmiller/nn-fap}.
The model, weights and code can also be found at  ~\href{https://www.nialljmiller.com/projects/FAP/FAP.html}{https://www.nialljmiller.com/projects/FAP/FAP.html}.
The training data used will be available upon any reasonable request to the first author. 



\bibliographystyle{mnras}
\bibliography{main_paper.bib} % if your bibtex file is called example.bib

%\begin{thebibliography}{99}
%\end{thebibliography}

%%%%%%%%%%%%%%%%%%%%%%%%%%%%%%%%%%%%%%%%%%%%%%%%%%

%%%%%%%%%%%%%%%%% APPENDICES %%%%%%%%%%%%%%%%%%%%%

\appendix

%%%%%%%%%%%%%%%%%%%%%%%%%%%%%%%%%%%%%%%%%%%%%%%%%%

% Don't change these lines
\bsp	% typesetting comment
\label{lastpage}
\end{document}